\documentclass[reprint,superscriptaddress,
 amsmath,amssymb,
 aps,prl,longbibliography,floatfix]{revtex4-1}

\usepackage{amsmath,amssymb}
\usepackage{graphicx}
\usepackage{color}
\usepackage{bm}
\usepackage{hyperref}
\hypersetup{
colorlinks=true,
urlcolor= blue,
citecolor=blue,
linkcolor= blue,
bookmarks=true,
}

\newcommand*\diff{\mathop{}\!\mathrm{d}}

\newcounter{para}
\newcommand\mypara{\par\refstepcounter{para}\noindent \textbf{\thepara}\indent}

\begin{document}

\title{Probing Image Potential States on the Topological Semimetal Antimony}

\author{Jian-Feng Ge}
\affiliation{Department of Physics, Harvard University, Cambridge, MA 02138, USA}

\author{Haimei Zhang}
\affiliation{School of Engineering and Applied Science, Harvard University, Cambridge, MA, 02138, USA}
\affiliation{Department of Physics, Wellesley College, Wellesley, MA 02481, USA}

\author{Yang He}
\affiliation{Department of Physics, Harvard University, Cambridge, MA 02138, USA}

\author{Zhihuai Zhu}
\affiliation{Department of Physics, Harvard University, Cambridge, MA 02138, USA}

\author{Yau Chuen Yam}
\affiliation{Department of Physics, Harvard University, Cambridge, MA 02138, USA}
\affiliation{Department of Physics and Astronomy, University of British Columbia, Vancouver, British Columbia V6T 1Z4, Canada}
\author{Pengcheng Chen}
\affiliation{Department of Physics, Harvard University, Cambridge, MA 02138, USA}
\author{Jennifer E. Hoffman}
\email[]{jhoffman@physics.harvard.edu}
\affiliation{Department of Physics, Harvard University, Cambridge, MA 02138, USA}
\affiliation{School of Engineering and Applied Science, Harvard University, Cambridge, MA, 02138, USA}

\date{\today}

\begin{abstract}
A point charge near the surface of a topological insulator (TI) with broken time-reversal symmetry is predicted to generate an image magnetic charge in addition to an image electric charge. We use scanning tunneling spectroscopy to study the image potential states (IPS) of the topological semimetal Sb(111) surface. We observe five IPS with discrete energy levels that are well described by a one-dimensional model. The spatial variation of the IPS energies and lifetimes near surface step edges shows the first local signature of resonant interband scattering between IPS, which suggests that image charges too may interact. Our work motivates the exploration of the TI surface geometry necessary to realize and manipulate a magnetic charge.
\end{abstract}






\maketitle 

\begin{figure}[tb]
\includegraphics{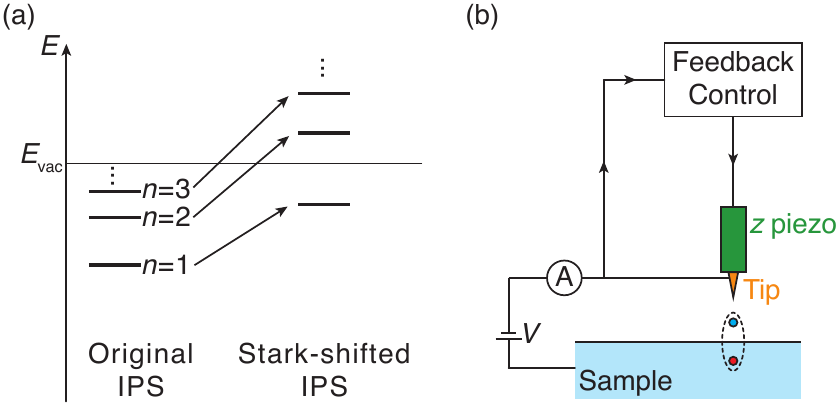}
\caption{Probing image potential states by STM. (a) Left: the Rydberg series of infinite discrete energy levels (left) approaches the vacuum  \(E_\mathrm{vac}\), as described by Eq.~\ref{eq:en}. Right: the quantized energy levels of image potential states can be shifted apart by stronger electric field (closer tip-sample distance, larger current), an effect known as the Stark shift. (b) A schematic drawing of the STM experimental setup. An electric image charge (red) is induced on a conducting sample when an external charge (blue) is placed above the surface. Together, these charges form a bound state (dashed oval). The measured tunneling current \(I\) is fed into a PID loop to adjust the tip-sample distance \(z\), by controlling the voltage applied on the \(z\) piezotube.}
\label{fig:cartoon}
\end{figure}

\mypara
The surface states of topological materials are spin-momentum locked, which reduces the local degrees of freedom and promotes an unusual electromagnetic response. In any material with conducting surface, when an electronic charge is located just outside the material, screening from surface charge will mimic an image electric charge \cite{Jackson1975textbook}. In a topological material, as in a normal metal, the attractive potential between the real and image charges can give rise to bound states known as image potential states (IPS) \cite{Echenique1991SurfSci}.  
However, in the topological material, the magnetic and electric degrees of freedom are additionally coupled by the topological magnetoelectric effect \cite{Qi2008PRB}. This effect opens a gap in the surface spectrum when time-reversal symmetry is broken, inducing a quantized Hall current and an effective magnetic monopole \cite{Qi2009Science}.
Thus the IPS in topological materials with broken time-reversal symmetry exhibit a combined image electric charge and image magnetic charge, which could be controlled by manipulating the real external electric charge near the surface. The manipulation of magnetic monopoles suggests possible applications such as circuitry based on magnetic charges \cite{Giblin2011Nphys}.
\mypara
Understanding the image electrical charge in topological materials is necessary to realize and control the image magnetic charge. The Coulomb potential between the external electric charge and its image results in a Rydberg series of energy levels below the vacuum energy level \(E_\mathrm{vac}\)
\begin{equation}
    E_n = E_\mathrm{vac} - \frac{0.85~\mathrm{eV}}{(n+a)^2},~ n=1, 2, ...\,,
    \label{eq:en}
\end{equation}
where \(a\) is a correction factor of the crystal field \cite{Echenique2002JESRP}. Moreover, a strong electric field can alter \(E_n\) due to the Stark effect \cite{Crampin2005PRL}. As a result, \(E_n\) is no longer a converging series below \(E_\mathrm{vac}\), but the levels shift to higher energy and spread farther apart, as shown in Fig.~\ref{fig:cartoon}(a).
\mypara
IPS have been well studied by \(z(V)\) spectroscopic measurements using scanning tunneling microscopy (STM) on metal surfaces \cite{Binnig1985PRL, Wahl2003PRL, Pivetta2005PRB, Dougherty2007PRB}, but IPS on topological materials were observed only recently by photoemission \cite{Niesner2012PRB, SobotaPRL2012, SobotaPRL2013, NiesnerPRB2014, NiesnerJESRP2014, Reimann2014PRB, Datzer2017PRB} and force microscopy \cite{Yildiz2019NM},
and have not yet been extensively characterized with local spectroscopy. Here we use spectroscopic STM to investigate IPS on the (111) surface of the topological semimetal Sb. Because of the typical electric field on order 1 V/nm applied between the tip and sample, the IPS are all Stark-shifted. We use \(z(V)\) spectroscopy to characterize the interaction of the external charge with the topological surface states, by quantifying the change in Stark shift with tip-sample junction setup conditions. Then we show how IPS from distinct terraces evolve across a step edge, and we demonstrate the first localized signature of resonant interband scattering. Our observations suggest a route to engineering interactions between images charges in topological materials.

\begin{figure}[tb]
\includegraphics{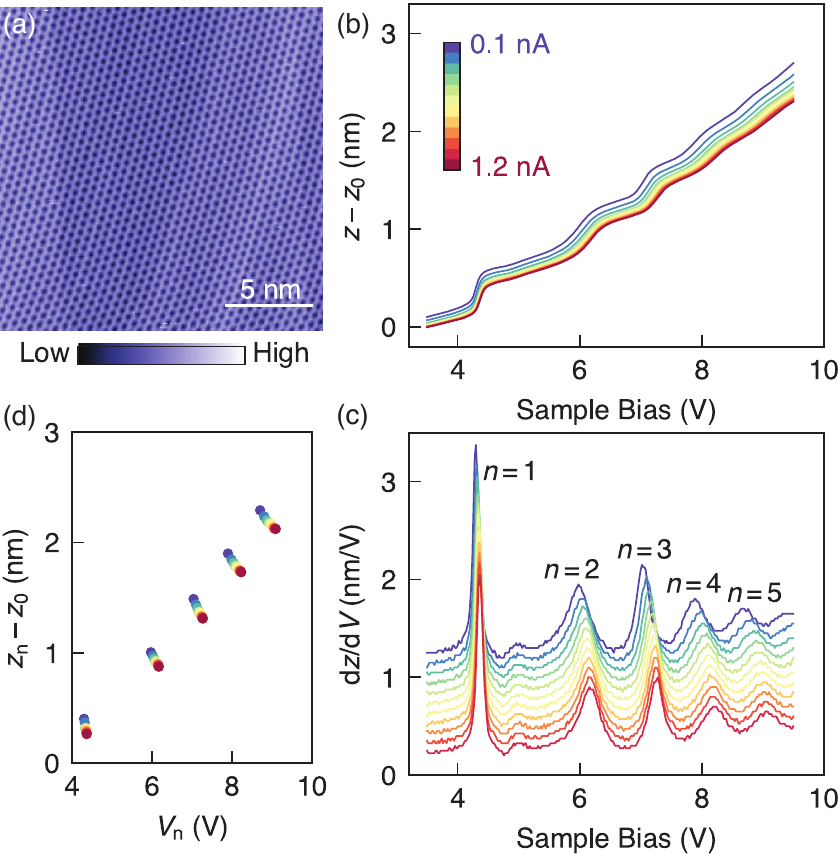}
\caption{Image potential states measured on Sb(111). (a) Topographic image of Sb(111) surface. Setup conditions: \(V_\mathrm{s} =\) 0.3 V, \(I_\mathrm{s} =\) 0.3 nA. (b) \(z(V)\) spectra acquired consecutively with the same tip and at the same location on Sb(111) at different current setpoints spaced by 0.1 nA intervals from 0.1 nA to 1.2 nA. Sample bias voltage \(V\) is swept from 3.5 V to 9.5 V. 
The measured tip-sample distance is relative to the setup height \(z_0\) at \(V_\mathrm{s} =\) 3.5 V, \(I_\mathrm{s} =\) 1.2 nA. (c) Numerical derivative of the \(z(V)\) spectra in (b). Each \(\diff z/\diff V(V)\) curve is shifted vertically by 0.1 nm/V for clarity. The first five IPS levels are labeled with \(n=1\), 2, 3, 4 and 5. (d) Relative \(z_n\) as a function of peak voltages \(V_n\). \(V_n\) is the center of each peak in (c) from Lorentzian fit. The \(z_n-z_0\) values corresponding to \(V_n\) are then extracted from (b).}
\label{fig:pointspectra}
\end{figure}

\begin{figure}[tb]
\includegraphics{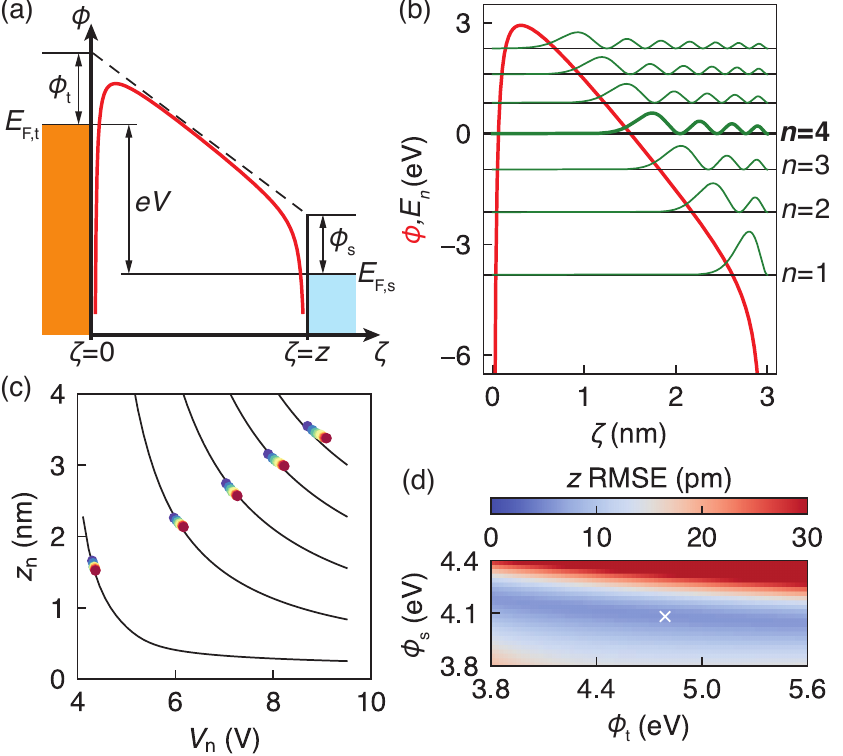}
\caption{Fitting IPS levels with a 1D model. (a) Energy diagram of the 1D model. \(E_\mathrm{F,t}\) and \(E_\mathrm{F,s}\) denote the Fermi energies of the tip and sample, respectively. The black dashed line and the red solid line show the the linear electrostatic potential of the STM and the full potential of the 1D model (Eq.~\ref{eq:phi}), respectively. (b) An example solution of the 1D model potential. The potential (red) is generated with the following parameters: \(V=\) 8.2 V, \(z =\) 3.0 nm, \(\phi_\mathrm{t}=\) 4.8 eV, \(\phi_\mathrm{s}=\) 4.1 eV. Normalized square moduli of eigenwavefunctions (green) are plotted on top of eigenenergy levels (black horizontal lines) for \(n\) from 1 to 7. Here the \(n=4\) eigenenergy coincides with \(E_\mathrm{F, t}\), which results in resonance tunneling. (c) Tip-sample distance \(z_n\) as a function of the peak voltage \(V_n\) of the best fit (black lines) using the 1D model. The experimental data (circles) are plotted with an offset \(z_0=\) 1.26 nm obtained from the best fit. (d) Fit residual map of the least-squares method for combinations of \(\phi_\mathrm{t}\) and \(\phi_\mathrm{s}\). The residual is presented as a root-mean-square error in \(z\) for all experimental data points in (c).  The cross mark denotes the best fit.}
\label{fig:1Dmodel}
\end{figure}

\begin{figure}[tb]
\includegraphics{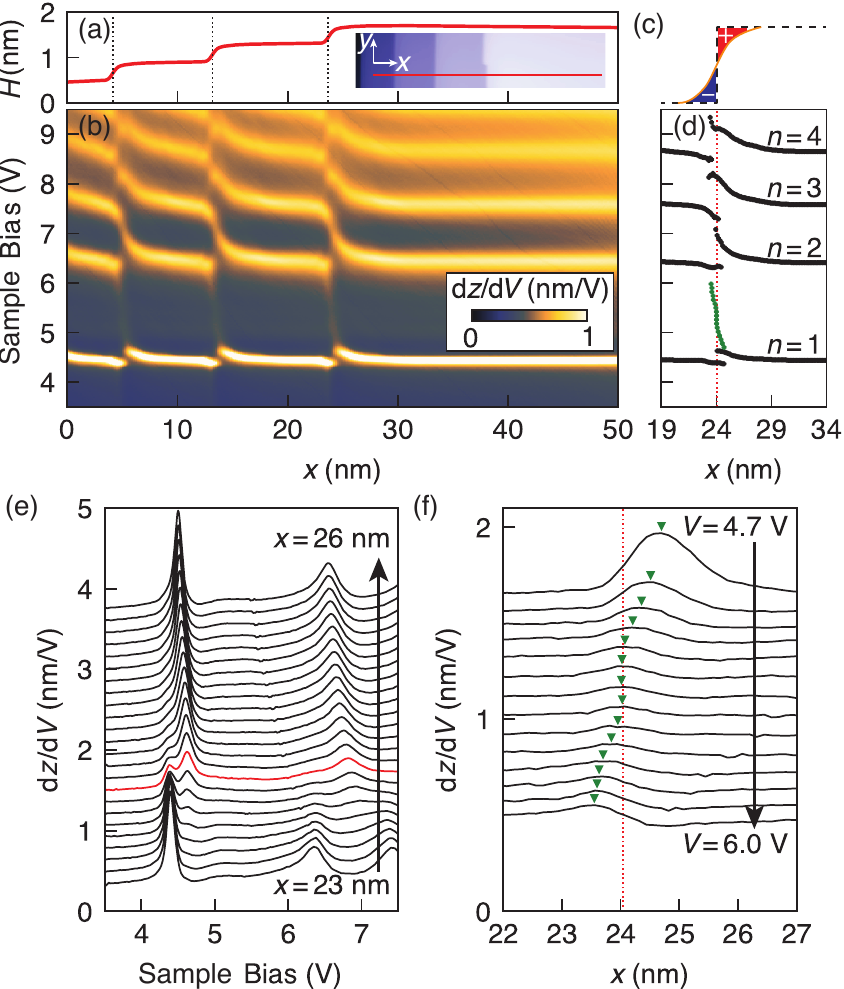}
\caption{Spatial dependence of IPS across surface steps. (a) Height profile (\(H\)) of the linecut. Dotted lines indicate positions of the steps. Inset: topographic image around the linecut (red).  Setup conditions: \(V_\mathrm{s}=\) 0.3 V, \(I_\mathrm{s}=\) 50 pA. (b) \(\diff z/\diff V(x, V)\) map along the line in (a). (c) Schematic of Smoluchowski effect showing the redistributed charge. Charge flows from the upper terrace to the lower terrace, resulting in net positive (red) and negative charge (blue) on two sides of the step (dashed lines), and a smooth equilibrium charge distribution (orange). (d) Peak voltages extracted (black) by Lorentzian fit of each \(\diff z/\diff V(V)\) spectrum along the line in (a). The green data points show the peak positions from Gaussian fit of \(\diff z/\diff V(x)\) to the transition state between the \(n=1\) and \(n=2\) IPS near a step edge. 
(e) \(\diff z/\diff V(V)\) spectra for \(x\) from 23 nm to 26 nm. The red curve denotes the spectrum taken on the step edge. Each curve is shifted by 0.1 nm/V for clarity. (f)  \(\diff z/\diff V(x)\) linecuts for bias voltage from 4.7 V to 6.0 V. Each curve is shifted by 0.15 nm/V for clarity. The centers of the Gaussian peaks are marked by the green triangles. The position of the step edge is denoted by the red dotted line in (d) and (f). 
}
\label{fig:stepdata}
\end{figure}

\mypara
Single crystals of Sb were cleaved in cryogenic ultra-high vacuum then loaded directly into our home-built STM. All measurements were carried out at 5 K with mechanically cut PtIr tips cleaned by field emission on Au. The observed (111) plane is atomically flat and free of defects in the 15 nm \(\times\) 15 nm area shown in Fig.~\ref{fig:pointspectra}(a).  
The experimental setup is illustrated in Fig.~\ref{fig:cartoon}(b). In \(z(V)\) spectroscopy, the STM operates in constant-current mode, where the feedback loop controls the tip-sample distance \(z\) to maintain a constant current \(I\). As the bias voltage \(V\) is increased, the tip is gradually pulled away from the sample.  When \(V\) is tuned in resonance with a Stark-shifted IPS level, there is an instantaneous increase in transmission probability and hence the current. This increase in current leads to an abrupt retraction of the tip, manifested in a \(z(V)\) spectrum as steps at resonance voltages \(V_n\).
\mypara
We searched for IPS in \(z(V)\) spectroscopic measurements on the Sb(111) surface with a set of tunneling currents ranging from 0.1 nA to 1.2 nA, as shown in Fig.~\ref{fig:pointspectra}(b). As the absolute tip-sample distance is unknown, we define a reference tip-sample distance \(z_0\) for the particular setup condition of \(V_\mathrm{s} =\) 3.5 V and \(I_\mathrm{s} =\) 1.2 nA, and we report the measured \(z\) consistently relative to the same \(z_0\). Each \(z(V)\) spectrum shows a series of steps at IPS resonance voltages superimposed on a gradually increasing background. We observe two shifts of the \(z(V)\) spectra as the current increases: the whole spectrum shifts down in the vertical axis; and the steps shift to the right in the horizontal axis. The shift in \(z\) can be understood as the tip being pushed towards the sample to achieve a higher current set point at the same bias voltage.  The shift in \(V\) is the consequence of the Stark effect: with an increasing current (and thus decreasing \(z\)), the electric field increases and shifts the IPS to higher energies and farther apart.
\mypara
In order to visualize IPS energy shifts more clearly, we numerically differentiate the \(z(V)\) spectra and plot the \(\diff z/\diff V(V)\) curves in Fig.~\ref{fig:pointspectra}(c). By counting the number of peaks, we observe five IPS up to 9.5 V. We fit each peak in Fig.~\ref{fig:pointspectra}(c) with a Lorentzian function and obtain \(V_n\) from the centers of the Lorentzian peaks. Comparing Figs.~\ref{fig:pointspectra}(b) and (c), we extract \(z_n\) relative to \(z_0\) at each peak voltage \(V_n\) and plot the values in Fig.~\ref{fig:pointspectra}(d). 
Fig.~\ref{fig:pointspectra}(d) shows that the Stark shift increases nonlinearly with increasing current and differs for each level index \(n\). We also note that the peak width generally increases at higher \(n\). The broadening of IPS peaks can be understood by elastic scattering of IPS electrons into the bulk continuum \cite{Pascual2007PRB}.
\mypara
To quantitatively understand the data in Fig.~\ref{fig:pointspectra}(d), we use a one-dimensional (1D) model to describe the electrical potential in the vacuum space between tip and sample, assuming that the radius of the tip is much larger than the absolute tip-sample distance \cite{Wahl2003PRL}. The 1D potential \(\phi\), as plotted in Fig.~\ref{fig:1Dmodel}(a), is the sum of the linear electrostatic potential from the bias of the STM tunnel junction, the image potential of the tip, and the image potential of the sample \cite{Pitarke1990SS}
\begin{equation}
    \phi(\zeta) = \phi_\mathrm{t} - (\phi_\mathrm{t} - \phi_\mathrm{s} + eV)\frac{\zeta}{z} - \frac{\alpha e^{2}}{4\pi\epsilon_0}\bigg(\frac{1}{\zeta}+\frac{1}{z-\zeta}\bigg).
    \label{eq:phi}
\end{equation}
In Eq.~\ref{eq:phi}, the variable \(\zeta\) is the 1D spatial coordinate, which has an origin at the surface of the tip (\(\zeta=0\)), \(\phi\) is expressed relative to Fermi level of the tip \(E_\mathrm{F,t}\), \(\alpha=1.15 \ln 2\) is a factor that accounts for all image charges \cite{Simmons1964JAP}, and \(\epsilon_0\) is the vacuum permittivity. The parameter \(V\) is the bias voltage applied to the sample with respect to tip. There are three unknown parameters: \(z\) is the absolute tip-sample distance, \(\phi_\mathrm{t}\) and \(\phi_\mathrm{s}\) are work functions of the tip and the sample, respectively.
We use the potential \(\phi (\zeta)\) to solve the 1D Schr{\"o}dinger equation numerically with the Numerov method \cite{Numerov1927ASNA} and find the resonance condition where \(E_n\) coincides with the Fermi level of the tip \(E_\mathrm{F,t}\), i.e. \(E_n = 0\). Figure \ref{fig:1Dmodel}(b) shows an example of seven derived eigenenergies and eigenwavefunctions with parameters \(V=\) 8.2 V, \(z =\) 3.0 nm. A resonance at the \(n=4\) IPS (\(E_4 = 0\)) indicates \(V_4=\) 8.2 V, \(z_4 =\) 3.0 nm, which can be compared to our data points in Fig.~\ref{fig:pointspectra}(d) with an adjustable parameter \(z_0\). We fit the data in Fig.~\ref{fig:pointspectra}(d) using the method of least squares for each pair of \(\phi_\mathrm{s}\) and \(\phi_\mathrm{t}\) in the grid shown in Fig.~\ref{fig:1Dmodel}(d), to minimize the root-mean-square error (RMSE) in \(z_n\) between all experimental data in Fig.~\ref{fig:pointspectra}(d) and the model result.  The best fit with minimal RMSE shown in Fig.~\ref{fig:1Dmodel}(c)  gives fit parameters \(\phi_\mathrm{t}=\) 4.79 eV, \(\phi_\mathrm{s}=\) 4.08 eV, and the offset \(z_0=\) 1.26 nm. Our \(\phi_\mathrm{t}\) and \(\phi_\mathrm{s}\) show reasonable agreement with the work functions of Au 5.1 eV (which likely coats the tip after field emission) and Sb 4.55-4.7 eV \cite{Michaelson1977JAP}, respectively.
\mypara
We next investigate the influence of surface defects on the IPS. Figure \ref{fig:stepdata} shows laterally resolved IPS along a line across three surface steps.
We observe bi-atomic-layer steps of height $\sim$4 \AA, as shown in Fig.~\ref{fig:stepdata}(a), consistent with previous reports \cite{Gomes2009Arxiv, Seo2010Nature, Yam2018Arxiv}. We acquire \(z(V)\) spectra at each point on a line [inset of Fig.~\ref{fig:stepdata}(a)], and plot the \(\diff z/\diff V(x, V)\) map in Fig.~\ref{fig:stepdata}(b). The energies of all the IPS peaks are constant far from the step edges but bend to higher (lower) energy near a step edge on the higher (lower) terrace. Despite the different terrace widths, the bipolar bending appears identical near all three step edges in \(\diff z/\diff V(x, V)\) map. Similar bipolar bending has been observed in nanostructures such as NaCl/Ag(100) \cite{Ploigt2007PRB}, Co/Au(111) \cite{Schouteden2009PRB}, Li/Cu(100) \cite{Stepanow2011PRB}, and defects on InAs(111) \cite{Martinez2015PRB}, and is attributed to the change of surface potential between different materials \cite{Ploigt2007PRB, Schouteden2009PRB, Zeljkovic2013PRB}. The fact that we see the bipolar bending of IPS on the elementary material Sb indicates a local variation of chemical potential near a step edge. This variation can be understood by the Smoluchowski effect \cite{Smoluchowski1941PR} as illustrated in Fig.~\ref{fig:stepdata}(c), where positive and negative charge builds up on the upper and lower edges of a step, respectively. The charge redistribution gives rise to a local dipole moment, which effectively acts as a lateral perturbation to the model potential in Eq.~\ref{eq:phi} \cite{Clinton1985PRB}. 

\begin{figure}[tb]
\includegraphics{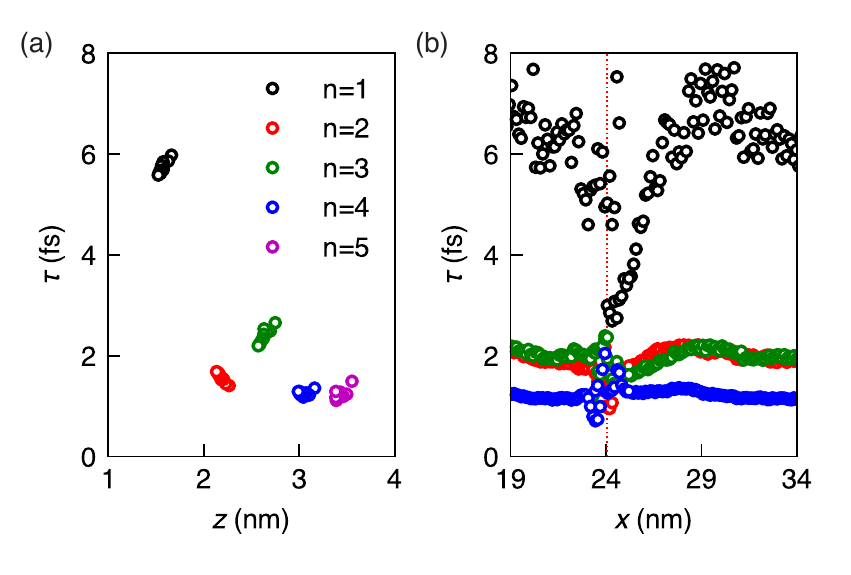}
\caption{Spatial dependence of the lifetimes \(\tau\) of IPS electrons. (a) Lifetime \(\tau\) as a function of tip-sample distance \(z\). Lifetime is extracted from Fig.~\ref{fig:pointspectra}(c), while \(z\) is estimated by \(z_n\) in Fig.~\ref{fig:1Dmodel}(c). (b) Lifetime \(\tau\) as a function of lateral distance \(x\) extracted from Fig.~\ref{fig:stepdata} near a step edge (red dotted line).}
\label{fig:lifetime}
\end{figure}

\mypara
We extract the IPS peak voltages in the same way as in Fig.~\ref{fig:pointspectra}, and plot these \(V_n\) as a function of distance in the \(x\)-direction in Fig.~\ref{fig:stepdata}(d). The bipolar bending follows approximately an exponential decay in \(x\), with a decay length on the order of 1 nm [see Fig.~\ref{fig:decaylength} in Appendix]. We also notice near the step edge each IPS peak splits into two, as exemplified in Fig.~\ref{fig:stepdata}(e). Although the bipolar bending of IPS has been observed, this splitting of IPS near a step edge has not been clearly characterized before. We speculate that this splitting may stem from the Stark effect caused by the additional local dipole moment or spilling of IPS electrons from one side of the step to the other \cite{Schouteden2009PRB}. 
\mypara
More interestingly, we observed an extra peak that departs from the \(n+1\) state on the lower terrace towards the \(n\) state on the higher terrace. This transition peak is most obvious for \(n=1\) in Fig.~\ref{fig:stepdata}(b), and it is localized within \(\sim\) 2 nm across the step as shown by green markers in Fig.~\ref{fig:stepdata}(d) and Fig.~\ref{fig:stepdata}(f). For higher \(n\), due to the increased width of the IPS peaks, it is difficult to distinguish between the transition peaks and the bending IPS. This transition peak from \(n+1\) to \(n\), which appears only in the upstairs direction, is a clear signature of resonant interband scattering \cite{Fauster2007PSS}. The directional preference in resonant interband scattering was previously noted in photoemission experiments on the stepped Cu(119) surface \cite{Roth2002PRL}. This cross-step scattering between distinct IPS states may provide a means to control the interaction between the expected induced magnetic charges when a magnetic layer is deposited on the stepped surface of a topological material \cite{Qi2009Science}.
\mypara
We now discuss the spatial evolution of the decay rate of the image charges.  The lifetime \(\tau\) of IPS can be estimated by \(\tau = \hbar/\Gamma\), where \(\Gamma\) is the full width at half maximum of the IPS peak \cite{Echenique2004SSR}. From the data in Fig.~\ref{fig:pointspectra}(c) and Fig.~\ref{fig:1Dmodel}(c) acquired far from a step edge, we plot \(\tau\) as a function of tip-sample distance \(z\) in Fig.~\ref{fig:lifetime}(a). The \(n=1\) IPS electrons live longer than those with higher \(n\), as the peak broadening due to inelastic decay increases dramatically for energies above \(E_\mathrm{vac}\) \cite{Crampin2006SurfSci}. We note that with increasing tip-sample distance, IPS lifetime increases for \(n=1\) and \(n=3\), but decreases for \(n=2\).  Compared to \(n=1\) IPS lifetime of \(17 \pm 4\) fs on topological insulator SnSb\(_2\)Te\(_4\) \cite{NiesnerPRB2014}, and \(\sim\)10-15 fs on Bi\(_2\)Te\(_3\) \cite{Yildiz2019NM}, the short \(n=1\) IPS lifetime \(\sim\)6 fs on Sb(111) can be attributed to the vanishing negative band gap in semimetal Sb and high availability of bulk states for IPS to scatter elastically into \cite{NiesnerJESRP2014}. We also extract from Fig.~\ref{fig:stepdata} the lateral dependence of IPS lifetime across a step edge in Fig.~\ref{fig:lifetime}(b). For \(n=1\) IPS electrons, the lifetime \(\tau\) drops near the step edge by 15\% on the lower terrace but by as much as 50\% on the upper terrace. The sizable asymmetry of our observed IPS lifetime is reminiscent of the finding on conventional Cu(119) that electrons running upstairs live longer than electrons running downstairs; the similarity can be taken as another piece of evidence for step-induced resonant interband scattering \cite{Roth2002PRL}. 
\mypara
In summary, we observed the first five IPS up to 9.5 V on the topological semimetal Sb(111) surface. The Stark-shifted IPS levels show good quantitative agreement with a simple 1D model that has been used to describe IPS on conventional metals. Additionally, laterally resolved IPS across surface steps show bipolar bending of the IPS levels and the first local signature of resonant interband scattering between IPS. Our study of IPS enriches the understanding of interacting image electrical charges on surfaces of topological materials, and paves the way towards the study of the associated magnetic charges and their interactions.

\begin{acknowledgments}
Experiments were supported by National Science Foundation DMR-1410480, and data analysis was supported by the Science and Technology Center for Integrated Quantum Materials under NSF DMR-1231319. We thank D.\ R.\ Gardner and Y.\ S.\ Lee for providing Sb single crystals, and we thank D.\ Yildiz for helpful conversations.
\end{acknowledgments}

\appendix*
\section{Appendix: Details of spatial dependence of IPS near a step edge}
\begin{figure}[h!]
\includegraphics{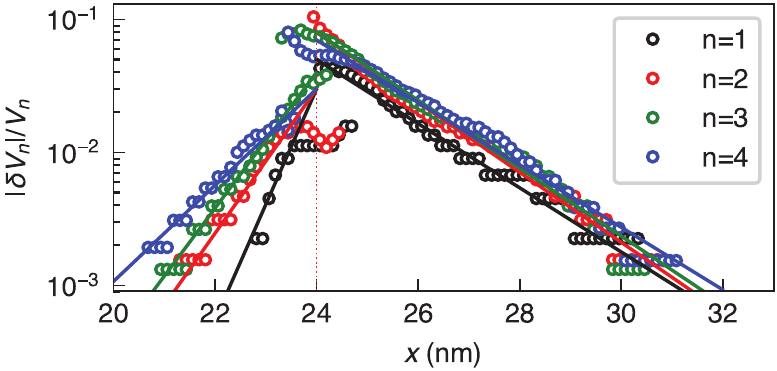}
\caption{Exponential decay of the bipolar bending near a step edge. Relative energy bending \(\lvert\delta V_n\rvert/V_n\) is calculated from Fig.~\ref{fig:stepdata}(d), where the (absolute) energy bending \(\lvert\delta V_n\rvert\) is the deviation from the average value of \(V_n\) far from the step edge. The solid curves are exponential fits to the data points for each \(n\) on the left and right side of the step edge.
}
\label{fig:decaylength}
\end{figure}

Here we extract the spatial dependence of energy bending \(\delta V_{n} (x) = V_{n} (x) - V_{n} (\infty)\), where we take average values of \(V_n\) for \(x=40\) to  \(x=50\) nm as \(V_{n} (\infty)\). We demonstrate in Fig.~\ref{fig:decaylength} the exponential decay of the relative energy bending \(\lvert\delta V_{n} \rvert / V_{n}\) as a function of \(x\) near the step edge at \(x=\) 24 nm. Note \(\delta V_{n}\) is negative on the left side of the step edge and its absolute value is displayed. We fit separately the data points on left and right sides with an exponential function \(\lvert\delta V_{n} (x) \rvert / V_{n} \propto \exp (-\lvert x-24\rvert/d_\mathrm{down, up})\), where \(d_\mathrm{down, up}\) is the decay length on the downstairs (left) or upstairs (right) side of the step edge. The fitted decay lengths for different \(n\) are summarized in Tab.~\ref{tab:1}. The decay length is also asymmetric with respect to the step. On the downstairs side of the step edge the the energy bending decays faster than that on the upstairs side by a factor of 2 to 3. This asymmetry may arise from the potential distribution across the step edge.
\begin{table}[h!]
    \centering
    \begin{tabular}{c | c | c}
        \hline
        \hline
             & \(d_\mathrm{down}\) (nm) & \(d_\mathrm{up}\) (nm) \\
        \hline
         \(n=1\) & 0.52 \(\pm\) 0.05 & 1.84 \(\pm\) 0.08 \\
         \(n=2\) & 0.84 \(\pm\) 0.04 & 1.84 \(\pm\) 0.05 \\
         \(n=3\) & 0.89 \(\pm\) 0.04 & 1.83 \(\pm\) 0.04 \\
         \(n=4\) & 1.18 \(\pm\) 0.04 & 1.91 \(\pm\) 0.05 \\
        \hline
        \hline
    \end{tabular}
    \caption{Decay length of bipolar energy bending extracted from fits (solid lines) in Fig.~\ref{fig:decaylength}.}
    \label{tab:1}
\end{table}

\bibliography{main}

\end{document}